\documentclass[conference]{IEEEtran}

\usepackage{cite}
\usepackage{amsmath,amssymb,amsfonts}
\usepackage{algorithmic}
\usepackage{graphicx}
\usepackage{textcomp}
\usepackage{xcolor}
\usepackage{multirow}
\usepackage{url}
\usepackage{hyperref}
\usepackage{booktabs}

\usepackage{threeparttable}
\usepackage{siunitx}

\usepackage{tabularx}

\usepackage{array}
\usepackage[table]{xcolor}
\usepackage{makecell}

\def\BibTeX{{\rm B\kern-.05em{\sc i\kern-.025em b}\kern-.08em
    T\kern-.1667em\lower.7ex\hbox{E}\kern-.125emX}}
\begin{document}

\title{Entropy-aware logistic regression for fusion of large-scale speaker recognition systems}


\author{\IEEEauthorblockN{Pierre-Michel Bousquet}
\IEEEauthorblockA{\textit{LIA, Avignon University, France}}
\and
\IEEEauthorblockN{Mickaël Rouvier}
\IEEEauthorblockA{\textit{LIA, Avignon University, France}}
}

\maketitle

\begin{abstract}
Score-level fusion based on logistic regression is widely used in speaker recognition to combine complementary systems.
However, conventional approaches assign fixed system-dependent coefficients and do not explicitly account for variations in the reliability of individual enrollment and test utterances.
Drawing on recent research on the entropy of deep learning-based speaker recognition models, this study incorporates an uncertainty component into the fusion process.
By exploiting both system-level complementarity and utterance-dependent uncertainty, the method achieves robust performance in large-scale speaker recognition tasks that involve highly variable characteristics of the speech signal.
These results demonstrate that model-entropy information provides a valuable complementary cue in large-scale scenarios.
\end{abstract}


\section{Introduction}
Large-scale speech recognition aims to address the challenge of developing speech recognition systems capable of processing a wide variety of speech signals and audio characteristics, including language, gender, emotional states, acoustic environments and devices, while meeting today’s high standards for robustness and efficiency.
In parallel with the production of giga-systems, which can potentially even be boosted by self-supervised learning, the multi-system approach remains an effective strategy for improving the accuracy of a speaker recognizer, and continues to attract significant research interest~\cite{Cumani2023,Cumani25_interspeech,Borgstrom2026}.
The term ``system'' refers to the combination of a neural network architecture, its configuration, and the training dataset used to optimize it.
Multi-system approaches exploit the complementarity between models trained with different architectures, configurations, or datasets, thereby improving the robustness and consistency of the resulting decisions.

Making the best use of scores generated by a set of more or less independent systems requires the development of score-level fusion methods that are as accurate and robust as possible.
Logistic regression remains one of the most widely used approaches, due to its statistical relevance and, notably, its inference capacity. The state-of-the-art method assigns a global coefficient to each set of scores from a given system (score-level fusion)
~\cite{brummer2010measuring,brummer2013likelihoodratiocalibrationusingpriorweighted}. Sparse fusion strategies have been proposed, based on correlation between score series, that remove some of them to enhance complementarity~\cite{Hautamaki_2013}.
The issue of taking into account specific conditions was addressed, with the aim of assigning coefficients at a finer level of granularity, down to the enrollment and test data for each comparison~\cite{Ferrer06,Cumani2023,Cumani25_interspeech,Borgstrom2026}.

Standard neural speaker recognition systems produce fixed-dimensional representations commonly referred to as x-vectors~\cite{Snyder2018}. Research has sought to link model entropy and x-vector uncertainty.
In~\cite{Lee2021}, it is proposed to connect them by integrating an auxiliary neural network, predicting uncertainty (linked to entropy) at the frame level. In this way, the Bayesian uncertainty modeling of the former i-vectors~\cite{Dehak09} is transposed into the deep architecture of x-vectors.
In~\cite{Silnova2020}, neural networks are trained to jointly estimate the x-vector and its uncertainty, with a particular focus on degraded or overly short audio signals.
Recently, specific loss functions (like \textit{Stochastic Variance Loss}) have been introduced to produce embeddings that are more robust to noise~\cite{Li2026xiuncertaintysupervisionrobust}.

In this work, we build upon previous research conducted on the entropy of deep learning-based speaker models~\cite{Bousquet2022}. This approach allows a degree of reliability to be assigned to each utterance representation produced by a system.
It is shown that the entropy of the model affects the x-vector representation, leading to uncertainty in the decision.
This allows the state-of-the-art logistic regression, which focuses on optimizing system-dependent coefficients, to be complemented by another logistic regression, this one attentive to the reliability of the speech material used in each voice comparison relative to each system.


This article is organized as follows: Section~\ref{sec:entropy} summarizes the initial approach to accounting for the system's entropy and presents the modifications made to this approach in this paper.
Section~\ref{sec:entropy_aware_logistic_regression} details the new logistic regression model that takes this entropy into account.
The experimental setup and results are presented in Sections~\ref{sec:experimental_protocol} and~\ref{sec:experiments_and_results}. We conclude in Section~\ref{sec:conclusion}.

\section{Entropy measures of a system}
\label{sec:entropy}

\begin{figure}[htbp]
\centerline{\includegraphics[scale=0.9]{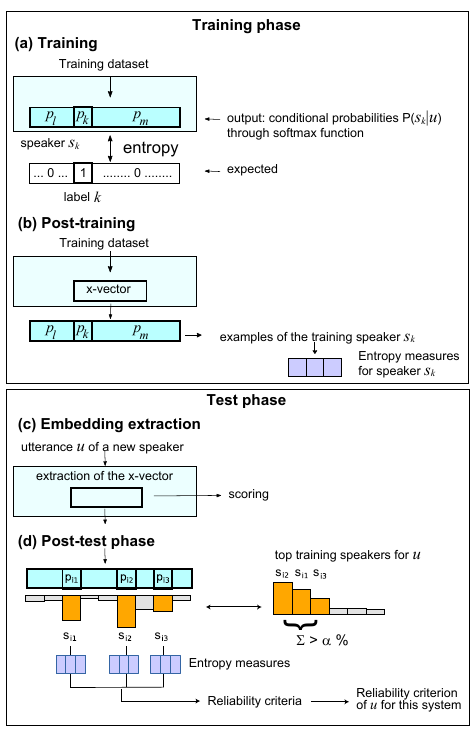}}
\caption{The additional procedures (panels (b) and (d)) inserted into the training and test phases (panels (a) and (c)), in order to extract entropy measures of a system and then use them as reliability criterion for each utterance of speakers unknown to the model.
}
\label{fig_entropy_measures}
\end{figure}
In~\cite{Bousquet2022}, a method is introduced to analytically assess the entropy of a speaker recognition (SR) system and to derive from it the uncertainty in the x-vector embedding.
For clarity, and because we have modified certain parts, the method is summarized in Figure \ref{fig_entropy_measures} and below.
In panel (a) of the figure, a model is trained, minimizing the entropy between conditional probabilities $p=\left\{P\left(s_{k}|u\right)  \right\}$ of training utterances $u$ from the training speakers $s_k$, estimated through a softmax function, and the expected target distribution. Once the training phase is achieved, a post-training procedure is carried out (in panel (b)): the x-vectors of the training data are extracted, then their conditional probabilities $p$ are used to determine entropy measures for each training speaker $s_k$.
The entropy measures proposed in~\cite{Bousquet2022} are defined as follows.
The first measure is the Kullback--Leibler divergence
$D_{KL}\left(\delta^{(k)} \| p\right)$ between the predicted distribution $p$ and the one-hot target distribution $\delta^{(k)}$, whose $k$th component is equal to $1$ and whose remaining components are equal to $0$. This divergence reduces to $-\log(p_k)$, which corresponds to the standard cross-entropy loss used during training. The aggregate measure for a given training speaker is obtained by averaging this quantity over all utterances from that speaker.

The second measure is a smoothing criterion inspired by the label-smoothing loss~\cite{Szegedy2015LabelSmoothing,Pereyra2017regularizing}. In this work, we replace Eq.~(3) of~\cite{Bousquet2022} with a symmetric Kullback--Leibler divergence. Using the same notation as in~\cite{Bousquet2022}, the measure is defined as follows:

\begin{align}
c\left(  \chi\right)    & =-D_{KL}\left(  u|\left[  p_{i}\right]  _{i\neq
y\left(  \chi\right)  }\right)  -D_{KL}\left(  \left[  p_{i}\right]  _{i\neq
y\left(  \chi\right)  }|u\right)  \nonumber\\
& =\frac{\sum_{i\neq y\left(  \chi\right)  }\log p_{i}}{N-1}-\frac
{\sum_{i\neq y\left(  \chi\right)  }p_{i}\log p_{i}}{1-p_{k}}+c%
\end{align}

The last measure is the dissimilarity between two training speakers in terms of entropy. With $s_k$ and $s_l$ having $n_k$ and $n_l$ examples:
\begin{equation}
\frac{1}{n_{_{k}}n_{_{l}}}\sum\limits_{p\in
s_{k}}\sum\limits_{q\in s_{l}}\left(  D_{KL}\left(  p||q\right)
+D_{KL}\left(  q||p\right)  \right)    
\end{equation}


In panel (c) of the figure~\ref{fig_entropy_measures}, the embeddings of new speakers of test set are extracted, and can be compared using cosine similarity or PLDA likelihoods~\cite{Prince07}.
In panel (d) of the figure, a post-extraction procedure is applied: the conditional probabilities of the test utterances $u$ are computed. The core of the method in \cite{Bousquet2022} is that this distribution is often concentrated in a few portions of the training population, called the \textit{top training speakers} of $u$.
Four reliability criteria for $u$ are computed, all based on the entropy of its top training speakers.
The first three are obtained by averaging the previous entropy measures on the top training speakers of this utterance.
The last one is the number of top-training speakers.
It should be noted that while too many top-speakers reveal the model's difficulty in fitting this utterance, a small number of top-speakers does not necessarily imply a lack of reliability.

The formulae for the final reliability criterion of an utterance given a system $\mathcal{S}$ differs from the one in \cite{Bousquet2022}. We use this formulae:
\begin{equation}
r\left(  u|\mathcal{S}\right)  =\frac{1}{4}%
{\displaystyle\sum\limits_{i=1}^{4}}
\dfrac{r_{i}\left(  u|\mathcal{S}\right)  -m_i}{M_i-m_i}%
\end{equation}
where $m$ and $M$ are minimal and maximal bounds of the criteria (the minimum and maximum, or very low and very high quantiles) computed on a single development set $D$ similar to the training set (no domain-specific development set as in~\cite{Bousquet2022}). This transformation leads to a final criterion in $[0,1]$ in increasing order of reliability.

To derive the reliability criteria for a comparison from those of its enrollment data $e$ and test data $t$, the formulae we propose is:
\begin{equation}
r\left(  e,t|\mathcal{S}\right)  =\min\left\{  r\left(  e|\mathcal{S}\right)  ,r\left(t|\mathcal{S}\right)  \right\}
\label{eqRelCrit}%
\end{equation}
The use of the minimum is justified below.

\begin{figure}[htbp]
\centerline{\includegraphics[scale=0.8]{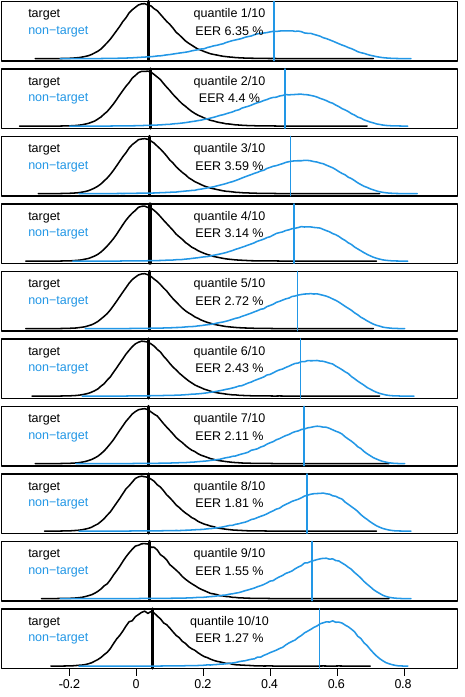}}
\caption{For the CommonBench evaluation: equal error rates and histograms of target/non-target scores per subsets of reliability quantile, from quantile 1, corresponding to the $10\%$ least reliable trials, to quantile 10, corresponding to the $10\%$ most reliable trials. The vertical lines are the averages of target/non-target scores. For the overall evaluation, the EER is equal to $3.46\%$.}
\label{fig_histos_tar_non}
\end{figure}

One could argue that model entropy is not a matter of concern, as it is well known that over-minimization of the total cross-entropy leads to overfitting and, consequently, to models that are incapable of performing inference on new data, especially data showing a significant mismatch with the training set.
However, this entropy could induce variable uncertainty in the point estimation of the embedding if the model is \textit{anisotropic} (i.e., possessing a locally variable goodness-of-fit).
This anisotropy is confirmed by the following analysis, reported in Figure~\ref{fig_histos_tar_non}: for the public CommonBench evaluation~\cite{hintz24_spsc}, the histograms of target (same speaker) and non-target (distinct speakers) scores are displayed, computed on one of the systems presented in Section~\ref{sec:experiments_and_results} (architecture: ResNet, training set: VoxCeleb2),  along with the equal error rates (EER). Each graph corresponds to a reliability-quantile subset, from the least reliable $10\%$ (top panel) to the most reliable $10\%$ (bottom panel). The EER is significantly correlated with reliability and ranges from $1.27\%$ to $6.35\%$ while the EER on the overall comparison set is equal to $3.46\%$.
The figure suggests further comments. Low-quality data do not necessarily lead to many errors. Rather, the reliability measure is significantly correlated with the risk of high error rate.
Furthermore, the graphs show that entropy induces distortion in the target scores and thus in within-speaker variability, while between-speaker variability remains approximately unchanged.

\begin{figure}[htbp]
\centerline{\includegraphics[scale=1.0]{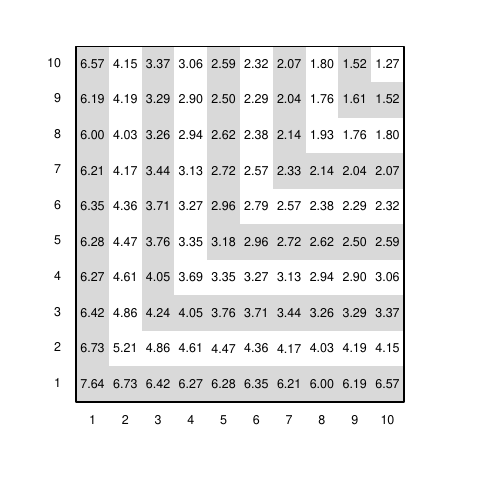}}
\vspace{-1cm}
\caption{EERs computed on subsets of CommonBench comparison trials per crossed deciles (enrollment, test) of reliability. Each gray or white stripe corresponds to the minimum of the enrollment and test deciles, confirming the relevance of equation (\ref{eqRelCrit}).}
\label{fig_EERs}
\end{figure}
The choice of the minimum in (\ref{eqRelCrit}) is justified by the analysis summarized in Figure~\ref{fig_EERs}.
With the same evaluation and configuration as above (CommonBench on ResNet and VoxCeleb2), the figure reports the EERs computed on subsets of comparison trials obtained by crossing given enrollment and test reliability deciles; for example, row 2, column 3, comparing enrollments in the second decile ($[10\%,20\%]$) and tests in the third decile ($[20\%,30\%]$) yields an EER of $4.86\%$.
It can be observed that the EERs in each gray or white stripe are relatively close. They correspond to the minimum of the enrollment and test deciles, confirming the relevance of (\ref{eqRelCrit}).

\section{Entropy-aware logistic regression}
\label{sec:entropy_aware_logistic_regression}
Let $\mathcal{S}=\left\{  \mathcal{S}_{i}\right\}$ denote a set of
$\left\vert \mathcal{S}\right\vert $ systems, and let $s_{i}\left(  e,t\right)$ denote the
score of the comparison between enrollment data $e$ and test data $t$ with the
$i^{th}$ system. Let $\theta_{tar}$ denote the \textit{target} hypothesis (i.e. same speaker).
Logistic regression (LR) for score fusion models the logit of the target event
as a linear combination of the scores, learned by maximum-likelihood estimation:

\begin{equation}
\text{logit~}P\left(  \theta_{tar}|e,t,\mathcal{S}\right)  =s_{\text{LR}%
}\left(  e,t\right)  =\beta_{0}+\sum\limits_{i=1}^{\left\vert \mathcal{S}%
\right\vert ~}\beta_{i}s_{i}\left(  e,t\right)
\end{equation}
Taking into account the results of the previous section, two sets of measures are now available for comparison: the scores and the entropic reliabilities. Their information can be merged by weighting the scores according to the reliability. For our purposes
$s_{i}\left(  e,t\right)  $ would become $s_{i}\left(  e,t\right)  r\left(e,t|\mathcal{S}_{i}\right)  /{\textstyle\sum\nolimits_{k=1}^{\left\vert \mathcal{S}\right\vert }}r\left(  e,t|\mathcal{S}_{k}\right)$. This raises several issues.
On the one hand, $r$ is a reliability indicator lying in $[0,1]$, while $s$ is a cosine metric in the embedding space or a probabilistic likelihood.
Weighting the latter based on the distribution of the former is purely empirical and not quantitatively justified. On the other hand, it is difficult to properly fit a continuous variable that is dependent on each individual observation.
To cope with this issue, a two-step strategy is employed: first, the initial criterion is replaced by a less stringent semi-parametric variable, then this variable is used as a generic input to a maximum-likelihood estimation procedure, such as logistic regression. The criterion $r$ is replaced by its quantile index: from 1 for the $10\%$ least reliable to 10 for the $10\%$ most reliable. 
Therefore, the reliability weight $w$ is an ordinal categorical variable between $1$ and $10$.
Given a set of scores to fuse, the corresponding weights $w$ are normalized to sum to 1: $w\rightarrow w/{\textstyle\sum}w$ and then a logistic regression is carried out on these weighted values. To implement LR with these weighted scores, they are scattered in a vector $\widehat{s}\in\mathbb{R}^{10}$ such that

\begin{equation}
\widehat{s}_{j}={\textstyle\sum\nolimits_{i/w\left(  e,t|\mathcal{S}_{i}\right)
=j}}\left(  \tfrac{w\left(  e,t|\mathcal{S}_{i}\right)}{
{\textstyle\sum\nolimits_{k=1}^{\left\vert S\right\vert }}
w\left(  e,t|\mathcal{S}_{k}\right)  }\right)s_{i}
\end{equation}

For example, with five scores $\left(  s_{1},...,s_{5}\right)$ whose reliability weights are, respectively, $\left(  8,1,5,5,4\right)  $, the scattered
vector $\widehat{s}$ will be equal to $\left[  \frac{1}{23}s_{2},0,0,\frac
{4}{23}s_{5},\frac{5}{23}s_{3}+\frac{5}{23}s_{4},0,0,\frac{8}{23}%
s_{1},0,0\right]  $.
Denoting the target/non-target ground-truth label by $l$,
the paired examples $(\widehat{s},l)$ can be used to fit a logistic regression model, fitting the coefficients $\left(  \gamma_{0},\left\{  \gamma
_{j}\right\}  _{j=1}^{10}\right)$ estimated by maximum likelihood.
The entropy-aware final score is equal to

\begin{multline}
s_{\text{e-LR}}\left(  e,t\right)=\\
\gamma_{0}+
{\textstyle\sum\limits_{j=1}^{10}}
\left(  \gamma_{j}%
{\textstyle\sum\limits_{\substack{i~/\\w\left(  e,t|\mathcal{S}%
_{i}\right)  =j}}}
\frac{w\left(  e,t|\mathcal{S}_{i}\right)  }{%
{\textstyle\sum\nolimits_{k=1}^{\left\vert \mathcal{S}\right\vert }}
w\left(e,t|\mathcal{S}_{k}\right)  }s_{i}\left(  e,t\right)
\right)
\end{multline}

which can be rewritten, to emphasize the scores $s_{i}\left(  e,t\right)$
\begin{equation}
\gamma_{0}+\sum\limits_{i=1}^{\left\vert \mathcal{S}\right\vert }\left(
\gamma_{w\left(  e,t|\mathcal{S}_{i}\right)  }\frac{w\left(
e,t|\mathcal{S}_{i}\right)  }{\sum\nolimits_{k=1}^{\left\vert
\mathcal{S}\right\vert }w\left(  e,t|\mathcal{S}_{k}\right)
}\right)  s_{i}\left(  e,t\right)
\end{equation}

While $\beta$ is system-dependent, $\gamma$ is utterance-reliability-dependent. Its purpose is to supplement the original LR score, not to compete with it. Therefore, the proposed final score is:
\begin{equation}
s_{\text{final}}\left(  e,t\right)  =\tfrac{1}{2}s_{\text{LR}}\left(
e,t\right)  +\tfrac{1}{2}s_{\text{e-LR}}\left(  e,t\right)
\end{equation}
The scores are simply equalized ($50\%,50\%$) to avoid over-tuning and provide a more realistic assessment of the method's performance.

\section{Neural Speaker Recognition Architectures and Training Protocol}
\label{sec:experimental_protocol}

This section describes the three main neural speaker verification architectures studied in this work, fwSE-ResNet, ECAPA2 and ReDimNet, and details the experimental protocol used for training. All models were trained using the Kiwano toolkit~\cite{rouvier2026}.

fwSE-ResNet-100 is a 2D-convolutional network utilizing a 100-layer ResNet backbone, organized into four residual stages comprising (3, 8, 18, 3) blocks and (128, 128, 256, 256) feature maps. Each block stacks three $3{\times}3$ convolutions using a BN $\rightarrow$ SiLU $\rightarrow$ Conv sequence to enhance nonlinearity. Frequency-wise Squeeze-and-Excitation (fwSE) modules are exclusively applied to the first two stages, enabling early channel-wise feature recalibration while keeping deeper layers computationally lightweight~\cite{rouvier2021studying}. Strided convolutions downsample time-frequency resolution, and the output is aggregated via Attentive Statistics Pooling into a 256-dimensional embedding. It processes 80-dimensional log-Mel filterbank inputs and uses an AM-Softmax loss optimized using cross-entropy.

ECAPA2~\cite{thienpondt2023ecapa2} is a hybrid network merging 1D and 2D convolutions through two components: a Local Feature Extractor (LFE) and a Global Feature Extractor (GFE). The LFE is a deep 2D-convolutional stack with five residual groups and fwSE modules, capturing fine-grained patterns across narrow frequencies. Strided convolutions downsample the frequency axis while preserving temporal resolution. This feature map is flattened and passed to the GFE, a 1D-convolutional sub-network featuring a Res2Net block. Channel-dependent attentive statistics pooling creates the 256-dimensional embedding. ECAPA2 accepts 256-dimensional spectrograms and employs Sub-Center Additive Angular Margin (AAM) Softmax loss.

ReDimNet~\cite{yakovlev24_interspeech} is another hybrid network that dynamically reshapes data between 1D and 2D formats to exploit both temporal and spectral-spatial processing. Following a 2D stem, intermediate representations share a 1D form. Successive stages reshape the signal into a 2D map for convolutions and ConvNeXt-like blocks, then revert it to 1D for an attention-based time-context block utilizing multi-head self-attention. Feature-map volume remains constant by offsetting frequency downsampling with proportional channel growth, enabling lossless cross-stage skip connections via softmax-weighted aggregations. The 15M-parameter B6 configuration uses Attentive Statistics Pooling to generate 256-dimensional embeddings from 72-dimensional inputs, trained with AM-Softmax loss.

Models and Training: all models generate 256-dimensional embeddings. They are trained for 51 epochs with a batch size of 512 and evaluated using cosine back-ends. AM-Softmax (margin=0.2, scale=30) is used for all models except ECAPA2, which uses AAM-Softmax.
All training data undergo 350-frame segmentation, followed by cepstral mean normalization and SpecAugment~\cite{Park_2019}. Additionally, either additive noise from MUSAN or reverberation based on simulated RIRs is applied per utterance (but not both simultaneously).

\section{Experiments and results}
\label{sec:experiments_and_results}

\begin{table}[htbp]
    \centering
    \caption{Corpora used for training and development}
    \label{tbl:datasets}

    \begin{threeparttable}
        \small
        \renewcommand{\arraystretch}{1.12}

        \begin{tabular}{
            @{}
            l
            l
            S[table-format=7.0, group-separator={\,}]
            S[table-format=5.0, group-separator={\,}]
            @{}
        }
            \toprule
            \textbf{Dataset}
            & \textbf{Language}
            & {\textbf{\#Utterances}}
            & {\textbf{\#Speakers}} \\
            \midrule

            CN-Celeb1-dev
            & Chinese
            & 39656
            & 642 \\

            CN-Celeb2-dev
            & Chinese
            & 382156
            & 1603 \\

            Common Voice\tnote{a}
            & Multiple
            & 507512
            & 5788 \\

            LibriSpeech
            & English
            & 214453
            & 1912 \\

            MLS\tnote{a}
            & European languages
            & 59199
            & 419 \\

            QASR-dev
            & Arabic
            & 192845
            & 1984 \\

            SdSV-dev
            & Persian
            & 62943
            & 472 \\

            VoxCeleb1
            & English\tnote{b}
            & 153516
            & 1251 \\

            VoxCeleb2
            & English\tnote{b}
            & 1063604
            & 5994 \\

            \midrule
            \textbf{Total}
            & {}
            & \bfseries 2675884
            & \bfseries 20065 \\
            \bottomrule
        \end{tabular}

        \begin{tablenotes}[flushleft]
            \footnotesize
            \item[a] Only non-English data are included.
            \item[b] Mainly English, including native and non-native speakers.
        \end{tablenotes}
    \end{threeparttable}
\end{table}

\begin{table}[htbp]
\caption{Training and development datasets\\used in our experiments}
\label{tbl:training_datasets}
\vspace{-0.3cm}
\center
\begin{tabular}[c]{lrr}
\toprule
\textbf{Training data}&\#utt.&\#spk\\
T1 CN-Celeb1-2+CommonVoice-non-English&929 324&8033\\
T2 LibriSpeech+MLS+QASR+SdSV&529 440&4787\\
T3 VoxCeleb2&1 063 604&5994\\\hline
\textbf{Development data}$^{(\mathrm{a})}$$^{(\mathrm{b})}$
&187 158&1500\\
development comparison dataset $^{(\mathrm{c})}$$^{(\mathrm{d})}$&\#non-target&\#target\\
&7 187 000&179 675\\
\bottomrule
\multicolumn{3}{l}{$^{(\mathrm{a})}$
\textit{Mixture of all corpora. The dataset proportions are the same as}}\\
\multicolumn{3}{l}{\textit{those in Table~\ref{tbl:datasets}}}\\
\multicolumn{3}{l}{$^{(\mathrm{b})}$
\textit{VoxCeleb2 is replaced by VoxCeleb1}}\\
\multicolumn{3}{l}{$^{(\mathrm{c})}$
\textit{Enrollment uses a single set of 5 utterances per speaker}}\\
\multicolumn{3}{l}{$^{(\mathrm{d})}$
\textit{for each test utterance: about 1 target and 40 non-target trials}}\\
\end{tabular}
\end{table}

We choose to take on the challenge of large-scale SR across a broad range of languages. Language is known to be a serious cause of mismatches in SR, and extending speaker recognition to various languages and cross-lingual conditions helps promote equity in speech sciences.
Other characteristics of the speech signal will be addressed in future work.
The corpora, training and development data used in our experiments are detailed in Tables~\ref{tbl:datasets} and~\ref{tbl:training_datasets}.
\subsection{On a homemade large-scale comparison dataset}











The method is first tested on a homemade comparison dataset. This dataset is intended to be as varied as possible in terms of language and to provide a sufficiently large number of trials, with a balance per test data. The enrollment data used to build a target model consist of a unique sample of 5 utterances per speaker. This condition is intended to reflect real-life SR tasks, in which enrollment includes all available information for each speaker, subject to a minimum-duration or minimum-number-of-segments requirement. 
The comparison dataset consists of 7 366 675 trials, including 179 675 “target” trials (same speaker), drawn from 240 771 utterances of 2000 speakers extracted from the corpora detailed in Table 1; these speakers are different from those used for training and development.
All score computations use the cosine metric. The enrollment x-vector is the $L_2$-normalized average of the $L_2$-normalized sample.

Results are reported in Table~\ref{tbl:results_test_trials} in terms of EER and of $C_{llr}$ (a more complete quality measure of score distribution~\cite{Brummer06}). Only the actual $C_{llr}$ is reported. No post-optimization (post-calibration or PAV algorithm~\cite{brummer2013pavalgorithmoptimizesbinary}) is performed.
The first part of the table reports results of the nine single systems obtained by pairing one of the three architectures with one of the three training datasets. These results allows comparisons of the efficiency and complementarity of the architectures.
The second part compares the logistic regression approaches on various system fusions. Note that, each time, the LR coefficients are estimated on the development set and applied to the test data.The addition of e-LR results in substantial performance improvements, according to both metrics, across all rows except the second one, which uses the ECAPA2 architecture. The last two rows correspond to the fusion of all systems, demonstrating the ability of this approach to combine systems.

\begin{table}[htbp]
\caption{Results of single systems and system fusions\\on the homemade large scale evaluation dataset of Table~\ref{tbl:training_datasets}$^{\text{(*)}}$}
\label{tbl:results_test_trials}
\vspace{-0.3cm}
\center
\begin{tabular}[c]{ccc|ccc|l|c|c}
\toprule
\multicolumn{3}{c|}{\textbf{Architecture}$^{\text{(**)}}$}&
\multicolumn{3}{c|}{\textbf{Training data}}&\makecell[c]{\textbf{Fusion}}&\textbf{EER}&\textbf{Cllr}\\
\textbf{A1}&\textbf{A2}&\textbf{A3}&\textbf{T1}&\textbf{T2}&\textbf{T3}&\makecell[c]{\textbf{method}}&\textbf{(\%)}&act\\\hline
$\checkmark$&&&$\checkmark$&&&--&2.47&0.807\\
$\checkmark$&&&&$\checkmark$&&--&2.95&0.821\\
$\checkmark$&&&&&$\checkmark$&--&2.32&0.805\\\hline

&$\checkmark$&&$\checkmark$&&&--&2.80&0.796\\
&$\checkmark$&&&$\checkmark$&&--&3.14&0.814\\
&$\checkmark$&&&&$\checkmark$&--&2.95&0.800\\\hline

&&$\checkmark$&$\checkmark$&&&--&2.88&0.800\\
&&$\checkmark$&&$\checkmark$&&--&3.28&0.812\\
&&$\checkmark$&&&$\checkmark$&--&2.75&0.797\\\hline
\hline
\multirow[c]{2}{*}{$\checkmark$}
&&&
\multirow[c]{2}{*}{$\checkmark$}&
\multirow[c]{2}{*}{$\checkmark$}&
\multirow[c]{2}{*}{$\checkmark$}&
LR&1.72&0.157\\
&&&&&&LR+e-LR&1.55&0.145\\\hline

&
\multirow[c]{2}{*}{$\checkmark$}
&&
\multirow[c]{2}{*}{$\checkmark$}&
\multirow[c]{2}{*}{$\checkmark$}&
\multirow[c]{2}{*}{$\checkmark$}&
LR&1.87&0.174\\
&&&&&&LR+e-LR&1.92&0.181\\\hline

&&
\multirow[c]{2}{*}{$\checkmark$}
&
\multirow[c]{2}{*}{$\checkmark$}&
\multirow[c]{2}{*}{$\checkmark$}&
\multirow[c]{2}{*}{$\checkmark$}&
LR&1.96&0.176\\
&&&&&&LR+e-LR&1.80&0.166\\\hline

\multirow[c]{2}{*}{$\checkmark$}
&
\multirow[c]{2}{*}{$\checkmark$}
&
&
\multirow[c]{2}{*}{$\checkmark$}&
\multirow[c]{2}{*}{$\checkmark$}&
\multirow[c]{2}{*}{$\checkmark$}&
LR&1.63&0.152\\
&&&&&&LR+e-LR&1.51&0.144\\\hline

\multirow[c]{2}{*}{$\checkmark$}
&&
\multirow[c]{2}{*}{$\checkmark$}
&
\multirow[c]{2}{*}{$\checkmark$}&
\multirow[c]{2}{*}{$\checkmark$}&
\multirow[c]{2}{*}{$\checkmark$}&
LR&1.72&0.157\\
&&&&&&LR+e-LR&1.51&0.141\\\hline

&
\multirow[c]{2}{*}{$\checkmark$}
&
\multirow[c]{2}{*}{$\checkmark$}
&
\multirow[c]{2}{*}{$\checkmark$}&
\multirow[c]{2}{*}{$\checkmark$}&
\multirow[c]{2}{*}{$\checkmark$}&
LR&1.79&0.167\\
&&&&&&LR+e-LR&1.66&0.158\\\hline

\multirow[c]{2}{*}{$\checkmark$}
&
\multirow[c]{2}{*}{$\checkmark$}
&
\multirow[c]{2}{*}{$\checkmark$}
&
\multirow[c]{2}{*}{$\checkmark$}&
\multirow[c]{2}{*}{$\checkmark$}&
\multirow[c]{2}{*}{$\checkmark$}&
LR&1.60&0.150\\
&&&&&&LR+e-LR&1.44&0.138\\


\bottomrule
\end{tabular}
    \begin{tablenotes}[flushleft]
        \footnotesize
        \item $^{\text{(*)}}$ Enrollment: unique speaker sample of size 5.
    \item $^{\text{(**)}}$ A1: ResNet A2: ECAPA2 A3: ReDimNet
    \end{tablenotes}
\end{table}

\begin{table}[htbp]
\caption{Results of fusion of all the systems with enrollment\\samples of various sizes\\on the same evaluation than Table~\ref{tbl:results_test_trials}}
\label{tbl:variable_size}
\vspace{-0.3cm}
\center
\begin{tabular}[c]{c|l|c|c}
\toprule
\textbf{Enrollment}&\makecell[c]{\textbf{Fusion}}&{\textbf{EER}}&\textbf{Cllr}\\
\textbf{sample size}&\makecell[c]{\textbf{method}}&\textbf{(\%)}&act\\\hline
\multirow[c]{2}{*}{1}&
LR&3.71&0.304\\
&LR+e-LR&3.53&0.298\\\hline

\multirow[c]{2}{*}{3}&
LR&1.98&0.184\\
&LR+e-LR&1.79&0.171\\\hline

\multirow[c]{2}{*}{5}&
LR&1.60&0.150\\
&LR+e-LR&1.45&0.138\\\hline

\multirow[c]{2}{*}{10}&
LR&1.32&0.128\\
&LR+e-LR&1.18&0.117\\
\bottomrule
\end{tabular}
\end{table}

Table~\ref{tbl:variable_size} reports results of the same experimental framework with enrollment samples of varying sizes.
On the one hand, this helps to highlight the impact of the amount of enrollment data available on the estimation of the target speaker model.
On the other hand, a detailed observation of the gain provided by e-LR along the enrollment sample size shows that the larger the enrollment sample, the greater the gain provided by e-LR: making available robust speaker models drastically reduces the uncertainty of the decision due to entropy.

\subsection{On public evaluations}
The method is tested on several publicly available evaluations.
The development comparison set used for estimating LR coefficients is one of those used for Table \ref{tbl:variable_size} experiments, depending on the evaluation. No fine-tuning was performed on the target domain for these evaluations.

Task 2 of the Short duration Speaker Verification (SdSV) Challenge~\cite{zeinali20_interspeech} is a text-independent SR evaluation based on the DeepMine dataset~\cite{deepmine2018odyssey,deepmine2019asru}, comprised of utterances from native Persian speakers and some non-native English speakers. The enrollment speech for each model comprised 1 to 29 utterances, with $47\%$ of the speakers having 3 utterances of enrollment. This evaluation was chosen because the organizers provided only $588$ development speakers, some of whom we included in the training dataset T2.
The rows of Table~\ref{tbl:SdSV} report successive results of all the single systems, then the best fusion with one, two architectures and finally all architectures. The development set employed here is the one with enrollment size of 3.
A comparison between the first nine and the last rows highlights the effectiveness of the multi-system fusion approach: with only $592$ Persian speakers in the training sets ($120$ in T1 and $472$ in T2) and $57$ in the development set, fusion by LR reduces the EER from $3.08\%$ to $2.44\%$ when using the ResNet architecture alone, and to $2.00\%$ when using all three architectures.
The addition of e-LR yields a significant improvement in performance in terms of both evaluation metrics, complementing the state-of-the-art LR.
\begin{table}[htbp]
\caption{Results of single systems and system fusions\\on SdSV-eval Task 2 evaluation}
\label{tbl:SdSV}
\vspace{-0.3cm}
\center
\begin{tabular}[c]{ccc|ccc|l|c|c}
\hline
\toprule
\multicolumn{3}{c|}{\textbf{Architecture}}&
\multicolumn{3}{c|}{\textbf{Training data}}&\makecell[c]{\textbf{Fusion}}&\textbf{EER}&\textbf{Cllr}\\
\textbf{A1}&\textbf{A2}&\textbf{A3}&\textbf{T1}&\textbf{T2}&\textbf{T3}&\makecell[c]{\textbf{method}}&\textbf{(\%)}&act\\
\hline

$\checkmark$&&&$\checkmark$&&&--&3.54&0.887\\
$\checkmark$&&&&$\checkmark$&&--&3.08&0.851\\
$\checkmark$&&&&&$\checkmark$&--&3.19&0.872\\
\hline

&$\checkmark$&&$\checkmark$&&&--&3.75&0.888\\
&$\checkmark$&&&$\checkmark$&&--&3.23&0.832\\
&$\checkmark$&&&&$\checkmark$&--&4.73&0.886\\
\hline

&&$\checkmark$&$\checkmark$&&&--&4.00&0.890\\
&&$\checkmark$&&$\checkmark$&&--&3.17&0.845\\
&&$\checkmark$&&&$\checkmark$&--&3.65&0.873\\
\hline
\hline

\multirow[c]{2}{*}{$\checkmark$}
&
&
&
\multirow[c]{2}{*}{$\checkmark$}&
\multirow[c]{2}{*}{$\checkmark$}&
\multirow[c]{2}{*}{$\checkmark$}&
LR&2.44&0.099\\
&&&&&&LR+e-LR&2.20&0.094\\\hline

\multirow[c]{2}{*}{$\checkmark$}
&
\multirow[c]{2}{*}{$\checkmark$}
&
&
\multirow[c]{2}{*}{$\checkmark$}&
\multirow[c]{2}{*}{$\checkmark$}&
\multirow[c]{2}{*}{$\checkmark$}&
LR&2.02 &0.086\\
&&&&&&LR+e-LR&1.88&0.082\\\hline

\multirow[c]{2}{*}{$\checkmark$}
&
\multirow[c]{2}{*}{$\checkmark$}
&
\multirow[c]{2}{*}{$\checkmark$}
&
\multirow[c]{2}{*}{$\checkmark$}&
\multirow[c]{2}{*}{$\checkmark$}&
\multirow[c]{2}{*}{$\checkmark$}&
LR&2.00&0.085\\
&&&&&&LR+e-LR&1.84&0.080\\


\bottomrule
\end{tabular}
\end{table}

CommonBench~\cite{hintz24_spsc} is a a large-scale speaker verification benchmark covering 101 languages and 11,793 speakers from Mozilla Common Voice 16.0~\footnote{\url{https://commonvoice.mozilla.org/en/}}.
Experiments are carried out on the training datasets T2 and T3 only, since T1 overlaps with the CommonBench dataset. Consequently, no Common Voice data are included in either the training set or the development set.
Results in Table~\ref{tbl:CommonBench} are reported in the same order as for SdSV. Here also, LR benefits from the additional e-LR component, in terms of EER as well as $C_{llr}$.

\begin{table}[htbp]
\caption{Results of single systems and system fusions\\on CommonBench evaluation}
\label{tbl:CommonBench}
\vspace{-0.3cm}
\center
\begin{tabular}[c]{ccc|ccc|l|c|c}
\toprule
\multicolumn{3}{c|}{\textbf{Architecture}}&
\multicolumn{3}{c|}{\textbf{Training data}}&\makecell[c]{\textbf{Fusion}}&\textbf{EER}&\textbf{Cllr}\\
\textbf{A1}&\textbf{A2}&\textbf{A3}&\textbf{T1}&\textbf{T2}&\textbf{T3}&\makecell[c]{\textbf{method}}&\textbf{(\%)}&act\\
\hline

$\checkmark$&&&&$\checkmark$&&--&4.16&0.867\\
$\checkmark$&&&&&$\checkmark$&--&3.46&0.861\\
\hline

&$\checkmark$&&&$\checkmark$&&--&5.16&0.891\\
&$\checkmark$&&&&$\checkmark$&--&10.1&0.874\\
\hline

&&$\checkmark$&&$\checkmark$&&--&5.85&0.891\\
&&$\checkmark$&&&$\checkmark$&--&4.79&0.861\\
\hline
\hline

\multirow[c]{2}{*}{$\checkmark$}
&
&
&
&
\multirow[c]{2}{*}{$\checkmark$}&
\multirow[c]{2}{*}{$\checkmark$}&
LR&2.91&0.336\\
&&&&&&LR+e-LR&2.74&0.290\\\hline

\multirow[c]{2}{*}{$\checkmark$}
&
\multirow[c]{2}{*}{$\checkmark$}
&
&
&
\multirow[c]{2}{*}{$\checkmark$}&
\multirow[c]{2}{*}{$\checkmark$}&
LR&2.87&0.281\\
&&&&&&LR+e-LR&2.66&0.210\\\hline

\multirow[c]{2}{*}{$\checkmark$}
&
\multirow[c]{2}{*}{$\checkmark$}
&
\multirow[c]{2}{*}{$\checkmark$}
&
\multirow[c]{2}{*}{}&
\multirow[c]{2}{*}{$\checkmark$}&
\multirow[c]{2}{*}{$\checkmark$}&
LR&2.82&0.278\\
&&&&&&LR+e-LR&2.59&0.196\\


\bottomrule
\end{tabular}
\end{table}

The TidyVoice Challenge 2026
\footnote{\url{https://tidyvoice2026.github.io/}}
 is a cross-lingual speaker verification evaluation built on a multilingual dataset derived from Common Voice~\cite{farhadipour2026tidyvoice}.
The method is tested on the validation trial set~\cite{Farhadipour2026plan}, which comprises 12 M trials from $59443$ utterances of $808$ speakers spanning $39$ languages. In particular, it contains 2 M target trials with language mismatch.
Table~\ref{tbl:TidyVoices} presents the results of the baseline provided by the organizers, of single systems then of the fusion of all systems.
For the same reasons mentioned above, the training set T1 is omitted. The multi-system approach combined with LR once again demonstrates its effectiveness, and the entropy-aware LR results in a relative gain of about 8\% for both measures.

\begin{table}[htbp]
\caption{Results of single systems and system fusions\\on TidyVoice 2026 evaluation}
\label{tbl:TidyVoices}
\vspace{-0.3cm}
\center
\begin{tabular}[c]{ccc|ccc|l|c|c}
\toprule
\multicolumn{3}{c|}{\textbf{Architecture}}&
\multicolumn{3}{c|}{\textbf{Training data}}&\makecell[c]{\textbf{Fusion}}&\textbf{EER}&\textbf{Cllr}\\
\textbf{A1}&\textbf{A2}&\textbf{A3}&\textbf{T1}&\textbf{T2}&\textbf{T3}&\makecell[c]{\textbf{method}}&\textbf{(\%)}&act\\
\hline

\multicolumn{6}{l|}{Baseline (organizers)}&
--&3.07&\\\hline\hline
$\checkmark$&&&&$\checkmark$&&--&3.85&0.874\\
$\checkmark$&&&&&$\checkmark$&--&3.01&0.867\\
\hline

&$\checkmark$&&&$\checkmark$&&--&4.81&0.883\\
&$\checkmark$&&&&$\checkmark$&--&4.22&0.868\\
\hline

&&$\checkmark$&&$\checkmark$&&--&3.96&0.873\\
&&$\checkmark$&&&$\checkmark$&--&3.42&0.859\\
\hline
\hline

\multirow[c]{2}{*}{$\checkmark$}
&

\multirow[c]{2}{*}{$\checkmark$}
&
\multirow[c]{2}{*}{$\checkmark$}
&
\multirow[c]{2}{*}{}&
\multirow[c]{2}{*}{$\checkmark$}&
\multirow[c]{2}{*}{$\checkmark$}&
LR&2.49&0.189\\
&&&&&&LR+e-LR&2.28&0.172\\


\bottomrule
\end{tabular}
\end{table}

\section{Conclusion}
\label{sec:conclusion}
In this study, an entropy-aware logistic regression is incorporated into score-level fusion for large-scale speaker recognition (SR). The results obtained from a large set of evaluation corpora demonstrate that incorporating this factor can improve the accuracy of the verification task.
More generally, this study reveals that the entropy of DNN-based SR models partially contributes to the uncertainty observed in SR tasks, and that this source of uncertainty can be quantified and incorporated into decision metrics or likelihoods.
Instead of competing with the state-of-the-art logistic regression, this approach is intended to complement the latter by providing information overlooked by the scoring methods, which rely on assumptions of a well-fitted and isotropic feature space.

\section{Acknowledgments}
This work was granted access to the HPC resources of IDRIS under the allocation 2026-AD011016065R1 made by GENCI.

\bibliographystyle{IEEEtran}
\bibliography{biblio_for_SLT_2026.bib}

\end{document}